\RequirePackage{fix-cm}
\documentclass[smallextended]{svjour3}       
\smartqed  
\usepackage{graphicx}
\usepackage{csquotes}
\usepackage[table]{xcolor}
\usepackage[authoryear]{natbib}
\usepackage{amsmath}
\usepackage{amssymb}
\begin{document}

\title{Common Counterfactual Belief of Rationality Subsumes Superrationality on Symmetric Games}
\titlerunning{CCBR Subsumes Superrationality on Symmetric Games}


\author{Ghislain Fourny}


\institute{G. Fourny \at
              ETH Z\"urich \\
              \email{ghislain.fourny@inf.ethz.ch}\\
}

\date{October 24, 2017}

\maketitle

\begin{abstract}
This paper shows that, for symmetric games in normal form, strategy profiles that satisfy Hofstadter's Superrationality criterion also satisfy both of Halpern's and Pass's criteria under Common Counterfactual Belief of Rationality: minimax-rationalizability and individual rationality.
\end{abstract}

\section{Introduction}

Most of the equilibrium concepts in game theory are Nash equilibria \citep{JNNCG}. This is because most of them are defined under the assumption that an agent considers the opponent's strategy to be counterfactually independent of their own strategy. In the case of two players playing a game in normal form, this means that, for example, the row player will hold a column for fixed, and pick the row that provides the best payoff in that column.

Over the last few decades though, there has been a few results that investigate what happens if this assumption does not hold, in other words, for agents that consider that, if they picked a different strategy, the opponent would also pick a different strategy. This is similar to the idea behind the Stackelberg equilibrium \citep{STACKELBERG} where the reaction of the opponent is anticipated and taken into account in the reasoning, but without the time dimension.

The idea of counterfactuals is not new and dates back to 1968 and 1973 with Robert \cite{Stalnaker1968} and David \cite{Lewis1973}. In the same period, Robert \cite{Nozick1969}, and then Martin \cite{NP} published Newcomb's problem, which materializes the concept of predicting the actions of a free, rational agent to the actual contents of an opaque box, and points out that a careful distinction between causal relationships and counterfactual dependences is paramount. The two-boxer resolution of Newcomb's problem is in agreement with the Nash paradigm---the choice of an agent is independent from the past---while the one-boxer resolution would correspond to the alternate, non-Nashian paradigm on which this paper focuses.

Douglas \cite{Hofstadter1983} defined the concept of superrational players  in a symmetric game such as the prisoner's dilemma. The idea is that a superrational player playing against another superrational player will consider that both will pick the same strategy, include this knowledge in her reasoning, and only reason on the diagonal of the normal-form matrix. In other words, a superrational player considers that her actions and the opponent's actions are counterfactually interdependent. As a consequence, the equilibrium reached will not (always) be a Nash equilibrium. Interestingly, after Douglas Hofstadter submitted his 20-way prisoner's dilemma challenge to his colleagues, Martin Gardner compared it to Newcomb's paradox in his explanations, but eventually chose to defect rather than cooperate, ``with a sign of regret.''

Jean-Pierre  \cite{Dupuy1992} showed a strong relationship between Newcomb's paradox and the prisoner's dilemma, confirming Martin Gardner's intuition, and, conjectured \citep{JPDPFNCE} that it should be possible to define an equilibrium concept for extensive form games that would implement the idea of counterfactual dependence between an agent's action and its prediction. Such an equilibrium would have an underlying notion of preemption to solve Grandfather's paradoxes that necessarily arise in the presence of a timeline, leading players to follow a Kantian imperative to avoid such paradoxes. It was proved in 2004 that this equilibrium exists, is unique and is Pareto optimal. The equilibrium and proofs were made public in 2014 as the Perfect Prediction Equilibrium (PPE) \citep{Fourny2014}.

An alternate course of research for counterfactual reasoning for games in extensive form is pursued by Richard \cite{Shiffrin2009}. Shiffrin's approach mainly differs from the PPE in that it disagrees on the relevance of preemption, and as such does not permanently discard outcomes deemed irrational in the middle of the reasoning.

\cite{Wolpert2011} formally established that both accounts of Newcomb's problem (one-boxer reasoning, two-boxer reasoning) can be explained with two different modelling of conditional probability distributions (incompatible Bayes nets), and as such, that both resolutions are acceptable.

Based on the work of David Lewis and Robert Stalnaker, \cite{Halpern:2013aa} defined the counterpart of Rationalizability for games in normal form in which players assume that their strategies are counterfactually interdependent, which they call Common Counterfactual Belief of Rationality (CCBR). Halpern and Pass define two criteria that strategy profiles can satisfy under the CCBR assumption: minimax rationalizability and individual rationality.

In this paper, we establish that, in the case of symmetric strategic games, superrational strategy profiles---we call them Hofstadter equilibria---satisfy both minimax rationalizability and individual rationality. In other words, the work of Joseph Halpern subsumes (is a superset of) Douglas Hofstadter's Superrationality. This formally establishes a link between CCBR and Superrationality.

\section{Background on non-Nashian game theory}

Before we come to this result, we give some background on games in normal form, Nash equilibria, Superrationality and CCBR.

\subsection{Games in normal form}

In game theory, games are typically expressed in two forms: normal form and extensive form. In extensive form, the game is expressed as a tree. At each node, a player picks a child node, and the leaves describe possible outcomes and are labelled with payoffs. On the other hand, in normal form, time plays no role and the payoffs are organized in a matrix---or a tensor of higher dimensionality than two if there are more than two players. Figure \ref{fig-normal-form} shows a two-player game in normal form. One player plays on the row, the other plays on the columns.

\begin{figure}
\begin{center}
\begin{tabular}{|r|c|c|}
\hline
& 1 & 2\\
\hline
A & $u_1(A, 1), u_2(A, 1)$ & $u_1(A, 2), u_2(A, 2)$\\
\hline
B & $u_1(B, 1), u_2(B, 1)$ & $u_1(B, 2), u_2(B, 2)$\\
\hline
\end{tabular}
\end{center}
\caption{A game in normal form, with two players that each can pick two strategies (A and B for the row player, 1 and 2 for the column player)}
\label{fig-normal-form}
\end{figure}

\begin{definition}[Game in normal form]

A game in normal form is defined with:

\begin{itemize}
\item a finite set of players $P$.
\item a set of strategies $\Sigma_i$ for each player $i\in P$.
\item a specification of payoffs $u_i(\overrightarrow\sigma)$ for each player $i\in P$ and strategy profile $\overrightarrow\sigma=(\sigma_j)_{j\in P}$.
\end{itemize}

\end{definition}

The payoff space only needs to be totally ordered (preference relation). In particular, when numbers are used, they are only meant as ordinals. In other words, comparing 1 to 1000 is no different than comparing 1 to 2.

Very often, given a player $i$, we denote as $\Sigma_{-i}$ the cartesian product of the remaining strategy spaces, and given a strategy profile $\overrightarrow\sigma$, we denote as $\sigma_{-i}$ the projection of the profile on $\Sigma_{-i}$. This very conveniently allows writing $u_i(\overrightarrow\sigma)$ as $u_i(\sigma_i, \sigma_{-i})$ with a slight abuse of notation that is always clear from the context.

Figures \ref{fig-prisoner-dilemma}, \ref{fig-chicken-game} and \ref{fig-coordination-game} show three famous examples of symmetric games. The payoffs have been normalized to the first natural numbers as cardinality is irrelevant.

\begin{figure}
\begin{center}
\begin{tabular}{|r|c|c|}
\hline
& Defect & Cooperate\\
\hline
Defect & 1, 1 & 3, 0\\
\hline
Cooperate & 0, 3 & 2, 2\\
\hline
\end{tabular}
\end{center}
\caption{The prisoner's dilemma. A player can either cooperate or defect. If both cooperate, they get more than if both defect. However, a player who unilaterally defects will get even more payoff than with mutual cooperation.}
\label{fig-prisoner-dilemma}
\end{figure}

\begin{figure}
\begin{center}
\begin{tabular}{|r|c|c|}
\hline
& Straight & Swerve\\
\hline
Straight & 0, 0 & 3, 1\\
\hline
Swerve & 1, 3 & 2, 2\\
\hline
\end{tabular}
\end{center}
\caption{The chicken game. A player can either stay straight or swerve. If both swerve, they get more (aka lose less) than if they both stay straight, and a player who unilaterally goes straight gets more payoff than if both swerve. The difference with the prisoner's dilemma, however, is that the "betrayed" player has interest in not reciprocating the betrayal (0 and 1 are swapped)}
\label{fig-chicken-game}
\end{figure}

\begin{figure}
\begin{center}
\begin{tabular}{|r|c|c|}
\hline
& Sushi & Pizza\\
\hline
Sushi & 1, 1& 0,0\\
\hline
Pizza & 0,0 & 2,2\\
\hline
\end{tabular}
\end{center}
\caption{The coordination game. In this game, the players have a mutual interest to pick the same strategy, even though one of the two strategies is better for both of them (aligned interest)}
\label{fig-coordination-game}
\end{figure}

\subsection{Nash equilibria}

Nash equilibria are defined having in mind that players hold their opponent's choices of strategies as fixed. The definition of a Nash equilibrium naturally arises as a strategy profile $\overrightarrow\sigma$ for which, for each player, the picked strategy $\sigma_i$ is the best response to the other players' strategies $\sigma_{-i}$.

Formally:

\begin{definition}[Nash equilibrium] Given a game $(P, \Sigma, u)$, a strategy profile $\overrightarrow\sigma$ is a Nash equilibrium if, for any player $i\in P$:

$$\forall \tau_i \in \Sigma_i, u_i(\sigma_i, \sigma_{-i}) \ge u_i(\tau_i, \sigma_{-i})$$
\end{definition}

Figures \ref{fig-prisoner-dilemma-nash}, \ref{fig-chicken-game-nash} and \ref{fig-coordination-game-nash} show the Nash equilibria for our example games.

\begin{figure}
\begin{center}
\begin{tabular}{|r|c|c|}
\hline
& Defect & Cooperate\\
\hline
Defect &  \cellcolor{black!25}1, 1 & 3, 0\\
\hline
Cooperate &  0, 3 &2, 2\\
\hline
\end{tabular}
\end{center}
\caption{The prisoner's dilemma. In the Nash equilibrium, both players defect.}
\label{fig-prisoner-dilemma-nash}
\end{figure}

\begin{figure}
\begin{center}
\begin{tabular}{|r|c|c|}
\hline
& Straight & Swerve\\
\hline
Straight & 0, 0 & \cellcolor{black!25}3, 1\\
\hline
Swerve & \cellcolor{black!25}1, 3 & 2, 2\\
\hline
\end{tabular}
\end{center}
\caption{The chicken game. It has two Nash equilibria: when players make opposite decisions.}
\label{fig-chicken-game-nash}
\end{figure}

\begin{figure}
\begin{center}
\begin{tabular}{|r|c|c|}
\hline
& Sushi & Pizza\\
\hline
Sushi & \cellcolor{black!25}1, 1& 0,0\\
\hline
Pizza & 0,0 & \cellcolor{black!25}2,2\\
\hline
\end{tabular}
\end{center}
\caption{The coordination game. All diagonal outcomes are Nash equilibria, i.e., players will not deviate if they made the same decision.}
\label{fig-coordination-game-nash}
\end{figure}

\bigskip

In contrast to the above Nash paradigm, the essence of Non-Nashian Game Theory, common to much of the work mentioned in this paper (Hofstadter, Fourny, Dupuy, Reiche, Shiffrin, Halpern, Pass), is that strategies are not optimized given a fixed opponent strategy, but the opponent strategy is rather assumed to be correlated to the strategy of the player at hand.

\subsection{Symmetric games}

There is a subclass of games that is of particular relevance, because players are interchangeable: symmetric games.

In a symmetric game, the strategy spaces are identical for all players, and the payoffs are defined in such a way that the game is invariant through a permutation of players.

\begin{definition}[symmetric game] A game is symmetric if

\begin{itemize}

\item the strategy spaces are identical

$$\forall i, j \in P, \Sigma_i = \Sigma_j = \Upsilon$$

\item the payoffs are symmetric

$$\forall \pi \in Sym(P), \forall \sigma \in \Sigma, u_i(\sigma) = u_{\pi(i)}(\sigma_{\pi(.)})$$

where $Sym(P)$ is the permutation group on P and $\sigma_{\pi(.)}$ denotes $(\sigma_{\pi(1)},$ $\sigma_{\pi(2)}, ..., \sigma_{\pi(|P|)})$.

\end{itemize}

\end{definition}

All games shown on Figures \ref{fig-prisoner-dilemma-nash}, \ref{fig-chicken-game-nash} and \ref{fig-coordination-game-nash} are symmetric.

\subsection{Superrationality}

Superrationality was introduced by Douglas Hofstadter in 1983 in a Scientific American column. In his own words:

\begin{displayquote}
``If reasoning dictates an answer, then everyone should independently come to that answer. Seeing this fact is itself the critical step in the reasoning toward the correct answer [...]. Once you realize this fact, then it dawns on you that \emph{either} all rational players will choose D \emph{or} all rational players will choose C. This is the crux.

\bigskip
 Any number of ideal rational thinkers faced with the same situation and undergoing similar throes of reasoning agony will necessarily come up with the identical answer eventually, so long as reasoning alone is the ultimate justification for their conclusion. Otherwise reasoning would be subjective, not objective as arithmetics is. A conclusion reached by reasoning would be a matter of preference, not of necessity.''
\end{displayquote}

It appears straightaway that the above consideration directly puts in question the fundamental assumption behind Nash equilibria that players consider their decisions to be independent from other players' decisions. Superrational players consider that their reasonings are correlated, not because of any causal effect or any kind of retrocausality, but because their reasonings and conclusions are identical.

The games described in the column are all symmetric, which is a requirement for the reasonings to be identical. Identical reasonings and conclusions mean that only outcomes on the diagonal of the normal form are considered. An equilibrium is reached if among all outcomes of the diagonal, it leads to the highest payoffs (which does not depend on the player as the game is symmetric).

Formally, this is expressed as follows.

\begin{definition}[Hofstadter equilibrium] Given a symmetric game in normal form, a strategy profile $\overrightarrow\sigma$ is an equilibrium reached by Superrational players (a Hofstadter equilibrium) if:

\begin{itemize}

\item the strategy profile is on the diagonal:

$$\exists\upsilon\in\Upsilon, \overrightarrow\sigma=(\upsilon, \upsilon, ..., \upsilon)$$

which we can also express as

$$\overrightarrow\sigma\in diag(\Sigma)$$

\item it maximizes the payoff on the diagonal

$$\forall \overrightarrow\tau \in diag(\Sigma), \forall i \in P, u_i(\overrightarrow\sigma) \ge u_i(\overrightarrow\tau)$$

\end{itemize}

\end{definition}

Figures \ref{fig-prisoner-dilemma-hofstadter}, \ref{fig-chicken-game-hofstadter} and \ref{fig-coordination-game-hofstadter} show the Hofstadter equilibria for our example games.

\begin{figure}
\begin{center}
\begin{tabular}{|r|c|c|}
\hline
& Defect & Cooperate\\
\hline
Defect & 1, 1 & 3, 0\\
\hline
Cooperate &  0, 3 & \cellcolor{black!25}2, 2\\
\hline
\end{tabular}
\end{center}
\caption{The prisoner's dilemma. Superrational players either both cooperate or both deviate. In a Hofstadter equilibrium, players both cooperate.}
\label{fig-prisoner-dilemma-hofstadter}
\end{figure}

\begin{figure}
\begin{center}
\begin{tabular}{|r|c|c|}
\hline
& Straight & Swerve\\
\hline
Straight & 0, 0 & 3, 1\\
\hline
Swerve & 1, 3 & \cellcolor{black!25}2, 2\\
\hline
\end{tabular}
\end{center}
\caption{The chicken game. Superrational players either both stay straight or swerve. In a Hofstadter equilibrium, players both swerve.}
\label{fig-chicken-game-hofstadter}
\end{figure}

\begin{figure}
\begin{center}
\begin{tabular}{|r|c|c|}
\hline
& Sushi & Pizza\\
\hline
Sushi & 1, 1& 0,0\\
\hline
Pizza & 0,0 & \cellcolor{black!25}2,2\\
\hline
\end{tabular}
\end{center}
\caption{The coordination game. Superrational players either both pick Sushi or Pizza. In a Hofstadter equilibrium, players both pick Pizza.}
\label{fig-coordination-game-hofstadter}
\end{figure}

\subsection{CCBR and minimax rationalizability}

A classical assumption made in game theory is Common Belief of Rationality: all players are rational, and believe that they all are, and believe that they all believe that they all are, and so on. In other words, rationality is \emph{profactually} assumed.

A framework in which players consider that their opponents' strategies are correlated to theirs, and who want to anticipate their reactions, needs to extend the assumption of rationality to other possible worlds: belief of rationality must also be \emph{counterfactually} assumed, recursively. This line of thought is common to the work by Hofstadter, Dupuy/Fourny/Reiche, Shiffrin, Halpern/Pass, etc\footnote{This is also to be put in relation with the terminology ``(common knowledge of) substantive rationality'' as discussed between Stalnaker, Aumann, Binmore and Halpern regarding the Backward Induction Paradox.}. Not only do players believe that they are rational, but they \emph{would} also believe so if they switched strategies. This is key to modelling the reaction of an opponent when considering alternate strategies.

Joseph Halpern and Rafael pass explicitly name---and formally define---this assumption as Common Counterfactual Belief of Rationality (CCBR): the beliefs of a player would also hold if this player would play a different strategy---they call this a counterfactual belief: all players are rational, and believe that they all are, and each one of them believes that all players would also be rational if they played otherwise, and so on. Building on this terminology, all players are rational, and counterfactually believe that they all are, and counterfactually believe that they all counterfactually believe that they all are, and so on.

Furthermore, CCBR does not assume that the opponents' strategies would be unchanged if a player unilaterally changed his strategy. Dropping this assumption leaves room for the above assumption on the counterfactuals. This is a bit similar to Stackelberg's idea, in that a player who changes his strategy would anticipate in his reasoning how other players would adapt. CCBR is thus a major change, away from the Nash paradigm.

Halpern and Pass further define minimax rationalizability to select strategies that make sense under CCBR. One way of characterizing minimax rationalizability is by iterated deletion of strategies that are minimax dominated.

Informally, in a game in normal form, a strategy of player $i$ is minimax dominated if there exists another strategy that guarantees him a higher payoff no matter what the opposite player does or would do. In other words, there is another strategy for which the minimum payoff is greater than the maximum payoff of the dominated strategy.

The intuition behind this definition is that, even if the player considers that the opponent's strategy is correlated with his choice, a minimax-dominated strategy will never be a good choice as the payoffs will nevertheless always be less, no matter what the assumed correlation is. If a player P picked a minimax-dominated strategy, even if the opponent's strategy is the best possible case for P, there is another strategy that would give him a higher payoff \emph{even if the (then possibly different) opponent's strategy were the worst possible case}.

This is to be put in contrast with ``classical'' rationalizability, in which	strategies which are not best responses are eliminated, which is a weaker requirement for elimination. In other words, a strategy that is minimax-dominated would also be eliminated according to classical rationalizability.

Figure \ref{fig-minimax-dominated} shows an example of game in which some strategies are minimax-dominated and can thus be eliminated under CCBR. The games shown on Figures \ref{fig-prisoner-dilemma}, \ref{fig-chicken-game} and \ref{fig-coordination-game} have no minimax-dominated strategies.

\begin{figure}
\begin{center}

\resizebox{\textwidth}{!}{
\begin{tabular}{lcccr}

\begin{tabular}{|r|c|c|c|}
\hline
& A & B & C\\
\hline
A & 9, 9& 8,6 & 5,1\\
\hline
B & 6,8 & 7,7 & 4,2\\
\hline
C & 1,5 & 2,4 & 3,3\\
\hline
\end{tabular}

&

$\Rightarrow$

&

\begin{tabular}{|r|c|c|c|}
\hline
& A & B & C\\
\hline
A & 9, 9& 8,6 & \cellcolor{black!25}5,1\\
\hline
B & 6,8 & 7,7 & \cellcolor{black!25}4,2\\
\hline
\rowcolor{black!25}C & 1,5 & 2,4 & 3,3\\
\hline
\end{tabular}

&

$\Rightarrow$

&

\begin{tabular}{|r|c|c|c|}
\hline
& A & B & C\\
\hline
A & 9, 9& \cellcolor{black!25}8,6 & \cellcolor{black!25}5,1\\
\hline
\rowcolor{black!25}B & 6,8 & 7,7 & 4,2\\
\hline
\rowcolor{black!25}C & 1,5 & 2,4 & 3,3\\
\hline
\end{tabular}

\end{tabular}
}
\end{center}
\caption{A symmetric 3x3 game in normal form, for which strategy C is minimax-dominated by both A and B for both players, and (after eliminating C) B is minimax-dominated by A for both players. The iteration goes from left to right, and eliminated profiles are marked in gray. C and B are not minimax-rationalizable, in other words, are not rational according to CCBR, even if each player considers the strategy of the other player not to be fixed}
\label{fig-minimax-dominated}
\end{figure}

\begin{definition}[minimax rationalizability]: Given a game $(P, \Sigma, u)$, given a player $i$, a strategy $\sigma_i$ is minimax-dominated\footnote{In the original paper \citep{Halpern:2013aa}, the opponent's strategy is taken from a subset of the opponent's strategies to account for successive eliminations. We are leaving out this aspect here for pedagogical reasons. Indeed, one can also mentally update $\Sigma$ in place as strategies get eliminated.} if

$$\exists \upsilon_i \in \Sigma_i, \min_{\tau_{-i}\in\Sigma_{-i}} u_i(\upsilon_i, \tau_{-i}) > \max_{\tau_{-i}\in\Sigma_{-i}} u_i(\sigma_i, \tau_{-i})$$

\end{definition}

Joseph Halpern and Rafael Pass give alternate characterizations of minimax-rationalizabi\-lity, but iterated deletion is the most intuitive one, and the one that we will use for our proof.

The concept of minimax-rationalizability, defined on strategies, extends to strategy profiles, i.e., a strategy profile can be considered to be minimax-rationalizable if all the strategies it is made of are minimax-rationalizable.

\subsection{CCBR and individual rationality}

Halpern and Pass define another concept under the assumption of CCBR: individual rationality of strategy profiles. The concept of individual rationality, unlike that of minimax-rationalizability, is not meant to be standalone, but rather to be a useful tool in some proofs related to games with translucent players\footnote{Offline discussion with Joseph Halpern at the ASIC conference in Interlaken, Switzerland, July 2017.}. Nevertheless, it is worth mentioning because superrationality, as we will see, also relates to individual rationality.

Informally, a strategy profile is individually rational if it Pareto-dominates a virtual strategy profile made of all ``best worst payoffs'', that is, each player gets at least what he has the power to guarantee himself by picking the strategy with the highest worst payoff, regardless of what the opponents do. Here again, one recognizes CCBR, because the opponent's strategy is not fixed, and instead, all possibilities are considered.

\begin{definition}[individually rational strategy profile] a strategy profile $\overrightarrow\sigma$ is individually rational if

$$\forall i \in P, u_i(\overrightarrow\sigma) \ge \max_{\tau_i\in\Sigma_i} \min_{\tau_{-i}\in\Sigma_{-i}} u_i(\overrightarrow\tau)$$

\end{definition}

Minimax-rationalizability and individual rationality, for strategy profiles, are not subsuming each other in any way: an individually rational strategy profile may not survive iterated minimax-deletion, and not all strategy profiles that survive it are individually rational. As Joseph Halpern and Rafael Pass point out, an individually rational strategy profile will always survive the first round of minimax elimination, but may get eliminated in the second. The intuitive reason is that, after a round of elimination, this strategy profile may ``lose'' its individual rationality because the elimination of some strategies can increase the threshold required for individual rationality.

\begin{figure}
\begin{center}
\begin{tabular}{|r|c|c|}
\hline
& Defect & Cooperate\\
\hline
Defect & \cellcolor{black!25}1, 1 & 3, 0\\
\hline
Cooperate &  0, 3 & \cellcolor{black!25}2, 2\\
\hline
\end{tabular}
\end{center}
\caption{The prisoner's dilemma with individually rational outcomes shown in gray. The virtual strategy profile with the best worst payoffs (maximins) is (1, 1). It would not be individually rational to have opposite strategies.}
\label{fig-prisoner-dilemma-individual}
\end{figure}

\begin{figure}
\begin{center}
\begin{tabular}{|r|c|c|}
\hline
& Straight & Swerve\\
\hline
Straight & 0, 0 & \cellcolor{black!25}3, 1\\
\hline
Swerve & \cellcolor{black!25}1, 3 & \cellcolor{black!25}2, 2\\
\hline
\end{tabular}
\end{center}
\caption{The chicken game with individually rational outcomes shown in gray. The virtual strategy profile with the best worst payoffs (maximins) is (1, 1). It would not be individually rational for both players to stay straight.}
\label{fig-chicken-game-individual}
\end{figure}

\begin{figure}
\begin{center}
\begin{tabular}{|r|c|c|}
\hline
& Sushi & Pizza\\
\hline
Sushi & \cellcolor{black!25}1, 1& \cellcolor{black!25}0,0\\
\hline
Pizza & \cellcolor{black!25}0,0 & \cellcolor{black!25}2,2\\
\hline
\end{tabular}
\end{center}
\caption{The coordination game with individually rational outcomes shown in gray The virtual strategy profile with the best worst payoffs (maximins) is (5, 5). All strategy profiles are individually rational.}
\label{fig-coordination-game-individual}
\end{figure}

\begin{figure}
\begin{center}
\begin{tabular}{|r|c|c|c|}
\hline
& A & B & C\\
\hline
A & \cellcolor{black!25}9, 9& \cellcolor{black!25}8,6 & 5,1\\
\hline
B &\cellcolor{black!25} 6,8 & \cellcolor{black!25}7,7 & 4,2\\
\hline
C & 1,5 & 2,4 & 3,3\\
\hline
\end{tabular}
\end{center}
\caption{The 3x3 game with individually rational outcomes shown in gray The virtual strategy profile with the best worst payoffs (maximins) is (5, 5). Indeed, strategy A guarantees for both players a minimum payoff of 5 regardless of what the other does.}
\label{fig-minimax-individual}
\end{figure}

Figures \ref{fig-prisoner-dilemma-individual} and \ref{fig-chicken-game-individual} show the individually rational strategies for the prisoner's dilemma and the chicken game. In these examples, some strategy profiles that are minimax rationalizable (they all are) are not individually rational (DC and CD are not, as well as Stay Straight-Stay Straight).

Figure \ref{fig-coordination-game-individual} shows the coordination game, where minimax rationalizability and individual rationality coincide (all strategy profiles qualify).

Figure \ref{fig-minimax-individual} shows the individually rational strategy profiles on the example we used for minimax rationalizability. Here, it can be seen that the strategy profiles AB, BA and BB are individually rational, but do not survive two rounds of minimax elimination.

\section{CCBR subsumes superrationality}

We now come to the main results of this paper. The theorems are straightforward to formulate given our introductory explanations.

\begin{theorem}
Given a symmetric game in normal form, a Hofstadter equilibrium is always minimax-rationalizable.
\end{theorem}

\begin{theorem}
Given a symmetric game in normal form, a Hofstadter equilibrium is always individually rational.
\end{theorem}

We now give the proofs of these two theorems.

\begin{proof}[Theorem 1]

In minimax rationalizability, the order of elimination is not relevant. Because of symmetry, if a strategy gets eliminated for a player, then it will be eliminated for all players. We reorder eliminations in such a way that strategies get eliminated for all players in batches, so that after each elimination, the game remains symmetric. We can now show that, for a symmetric game, a Hofstadter equilibrium cannot get minimax-eliminated.

Let $\overrightarrow\sigma$ be a Hofstadter equilibrium. We can write it as $\overrightarrow\sigma=(\sigma, \sigma, ..., \sigma)$ for some $\sigma\in\Upsilon$.

By definition of the maximum:

$$\max_{\tau_{-i}\in\Sigma_{-i}} u_i(\sigma, \tau_{-i}) \ge u_i(\sigma, \sigma, ..., \sigma)$$

because $(\sigma, ..., \sigma)$ is in the set over which the maximum is taken (opponents' strategies).

Let now $i$ denote a player, and $\upsilon\in\Upsilon$ now be any of its strategies.

By definition of a Hofstadter equilibrium, the payoffs are maximal on the diagonal, so that:

$$u_i(\sigma, \sigma, ..., \sigma) \ge u_i(\upsilon, \upsilon, ..., \upsilon)$$

Finally, by definition of the minimum:

$$u_i(\upsilon, \upsilon, ..., \upsilon) \ge \min_{\tau_{-i}\in\Sigma_{-i}} u_i(\upsilon, \tau_{-i}) $$

because $(\sigma, ..., \sigma)$ is in the set over which the maximum is taken (opponents' strategies).

By transitivity, we get:

$$\max_{\tau_{-i}\in\Sigma_{-i}} u_i(\sigma, \tau_{-i}) \ge \min_{\tau_{-i}\in\Sigma_{-i}} u_i(\upsilon, \tau_{-i})$$

which directly contradicts the existence of a strategy that allows minimax-domination, and this holds for any player.

There is one more thing to say for the proof to be complete. After an iteration of deletion of minimax-dominated strategies as described above, a Hofstadter equilibrium remains a Hofstadter equilibrium. This is because eliminating other rows or columns than that of the maximum diagonal payoff does not affect this maximum diagonal payoff. Hence, a Hofstadter equilibrium will recursively survive all iterations and, in the end, satisfy minimax rationalizability. $\square$
\end{proof}

\begin{proof}[Theorem 2]

Let $\overrightarrow\sigma$ be a Hofstadter equilibrium. We can write it as $\overrightarrow\sigma=(\sigma, \sigma, ..., \sigma)$ for some $\sigma\in\Upsilon$.
Let $i$ denote a player.

By definition of a Hofstadter equilibrium, the payoffs are maximal on the diagonal, so that:

$$u_i(\sigma, \sigma, ..., \sigma) \ge \max_{\upsilon\in\Upsilon} u_i(\upsilon, \upsilon, ..., \upsilon)$$

Furthermore, for any strategy $\upsilon$, 

$$u_i(\upsilon, \upsilon, ..., \upsilon) \ge \min_{\tau_{-i}\in\Sigma_{-i}}  u_i(\upsilon, \tau_{-i})$$

(the minimum payoff on its line can only be smaller than the payoff on the diagonal).

Combining the above inequalities:

$$u_i(\sigma, \sigma, ..., \sigma) \ge \max_{\upsilon\in\Upsilon} u_i(\upsilon, \upsilon, ..., \upsilon) \ge \max_{\upsilon\in\Upsilon} \min_{\tau_{-i}\in\Sigma_{-i}}  u_i(\upsilon, \tau_{-i})$$

Considering that $\upsilon$ is a mute variable and that $\Upsilon=\Sigma_i$ (symmetric game),

$$u_i(\sigma, \sigma, ..., \sigma) \ge  \max_{\tau_i\in\Sigma_i} \min_{\tau_{-i}\in\Sigma_{-i}}  u_i(\tau_i, \tau_{-i})$$

this fulfils the definition of an individually rational outcome. $\square$
\end{proof}

\begin{figure}
\resizebox{\textwidth}{!}{
\includegraphics{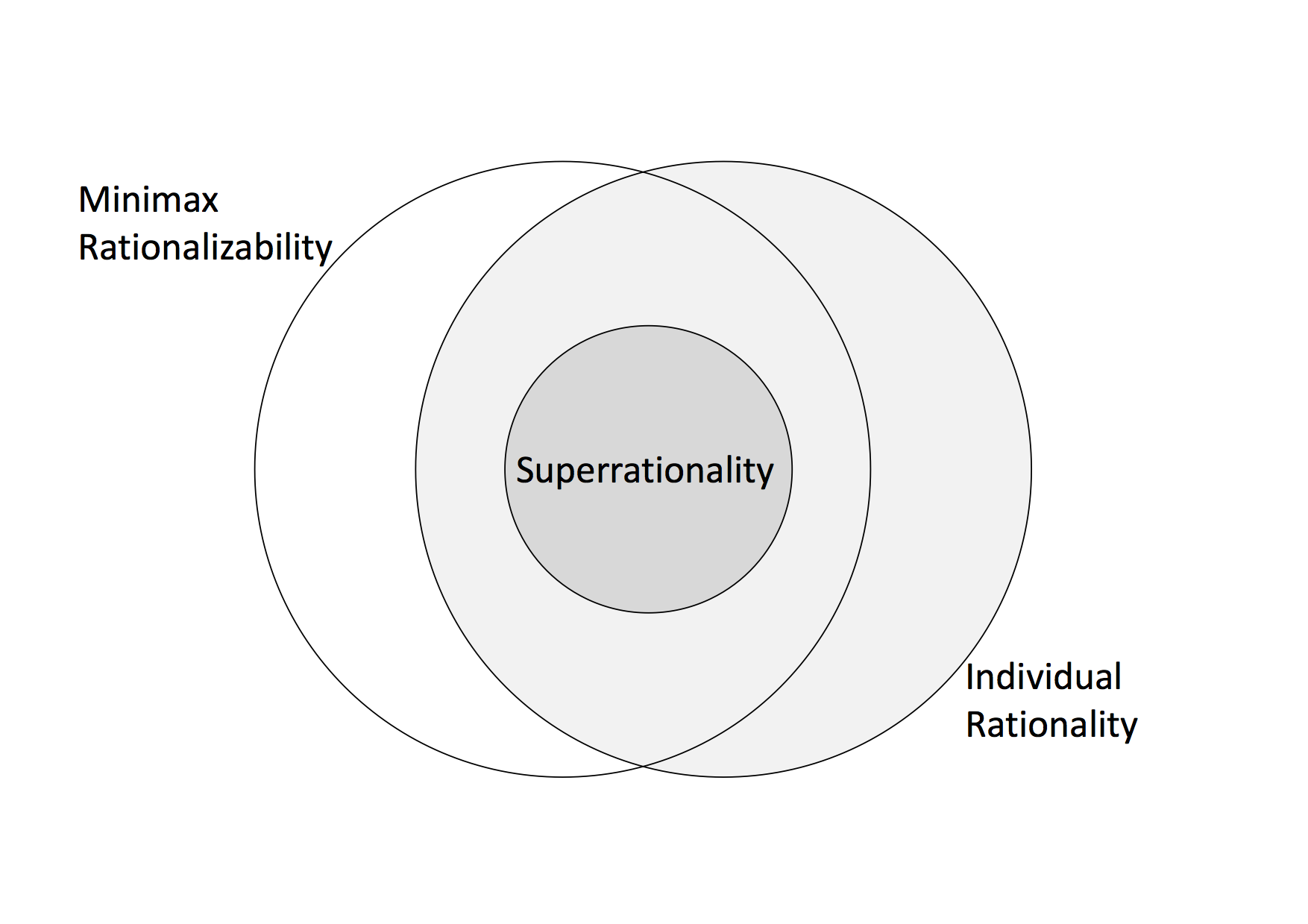}
}
\caption{A diagram depicting the relationship between minimax rationalizability, individual rationality, and Superrationality for strategy profiles.}
\label{fig-venn}
\end{figure}

Note that both inclusions are strict: there are minimax-rationalizable strategy profiles that are not Hofstadter equilibria (e.g., DC, CD, DD in the prisoner's dilemma), as well as individually rational strategy profiles that are not Hofstadter equilibria (e.g., Swerve-Stay Straight in the chicken game, which is not even on the diagonal, or Sushi-Sushi in the coordination game, which is not optimal).

\section{Conclusion and future work}

We have shown that Superrational players also behave in a way that is consistent with Joseph Halpern's and Rafael Pass's Common Counterfactual Belief of Rationality. This shows that Halpern's and Pass's work manages to capture the behavior described by Douglas Hofstadter. This validates the relevance of his rationalizability criteria for settings that deviate from Nashian game theory by assuming that a switch of strategy would be correlated with a switch of strategy by the opponents. The inclusions between the three concepts are depicted on Figure \ref{fig-venn}.

Further investigations include finding out whether the Perfect Prediction Equilibrium, as well as Shiffrin's equilibrium, are also subsumed by CCBR. Interestingly, it is straightforward to show with a counterexample that the PPE (with the game converted to its normal form equivalent) is not always individually rational, because CCBR expressed on the normal form cannot capture preemption. However, we conjecture that the PPE may qualify as minimax rationalizable. 

\section{Acknowledgement}

I am grateful to Joseph Halpern for providing feedback on a first draft of this paper, as well as to Jean-Pierre Dupuy, St\'ephane Reiche, Alexei Grinbaum, Bernard Walliser, Richard Shiffrin, Robert French for motivating and challenging discussions on the topic of Non-Nashian Game Theory.


\bibliographystyle{spbasic}      

\end{document}